\documentclass[reprint,superscriptaddress,amsmath,amssymb,aps,pra,floatfix]{revtex4-1}

\usepackage{amsfonts}
\usepackage{amsmath}
\usepackage{amsthm}
\usepackage{mathrsfs}
\usepackage{amscd}
\usepackage{amssymb}
\usepackage{subfigure}
\usepackage{amsxtra}
\usepackage{dcolumn}
\usepackage{bm}           
\usepackage{bbm}
\usepackage{graphicx}
\usepackage{epstopdf}

\usepackage[colorlinks]{hyperref}

\newcommand{\la}{\langle}
\newcommand{\ra}{\rangle}
\newcommand{\bra}[1]{\ensuremath{\langle#1|}}
\newcommand{\ket}[1]{\ensuremath{\left|#1\right\rangle}}

\newcommand{\ketbra}[2]{\ensuremath{\left| #1 \right\rangle \left\langle #2 \right|}}

\newcommand{\cL}{\mathcal{L}}
\newcommand{\cD}{\mathcal{D}}

\newcommand{\dg}{\dagger}
\newcommand{\da}{\dagger}

\newcommand{\Op}[1]{\hat{#1}}

\newcommand{\osigma}{\Op{\sigma}}

\newcommand{\oH}{\Op{H}}
\newcommand{\oA}{\Op{A}}

\newcommand{\oP}{\Op{P}}

\newcommand{\oD}{\Op{D}}
\newcommand{\oV}{\Op{V}}

\newcommand{\oa}{\Op{a}}
\newcommand{\ob}{\Op{b}}
\newcommand{\ox}{\Op{x}}
\newcommand{\otH}{\Op{\tilde{H}}}

\newcommand{\tr}{\ensuremath{{\rm tr}}}

\newcommand{\dd}{\mathrm{d}}

\begin{document}

\preprint{APS/123-QED}
\title{Maxwell's lesser demon: a quantum engine driven by pointer measurements}
\author{Stella Seah}
\affiliation{Department of Physics, National University of Singapore, 2 Science Drive 3, Singapore 117542, Singapore}

\author{Stefan Nimmrichter}
\affiliation{Max Planck Institute for the Science of Light, Staudtstra{\ss}e 2, 91058 Erlangen, Germany}

\author{Valerio Scarani}
\affiliation{Department of Physics, National University of Singapore, 2 Science Drive 3, Singapore 117542, Singapore}
\affiliation{Centre for Quantum Technologies, National University of Singapore, 3 Science Drive 2, Singapore 117543, Singapore}

\date{\today}
             
\begin{abstract}
We discuss a self-contained spin-boson model for a measurement-driven engine, in which a demon generates work from thermal excitations of a quantum spin via measurement and feedback control. Instead of granting it full direct access to the spin state and to Landauer's erasure strokes for optimal performance, we restrict this demon's action to pointer measurements, i.e.~random or continuous interrogations of a damped mechanical oscillator that assumes macroscopically distinct positions depending on the spin state. The engine can reach simultaneously the power and efficiency benchmarks and operate in temperature regimes where quantum Otto engines would fail. 
\end{abstract}
\maketitle

Conventionally, thermal machines operate through the interaction of a working medium with hot and cold reservoirs. In the context of quantum thermodynamics, interest has been raised in finding non-thermal resources such as coherence \cite{scully2003,scully2010,scully2011,uzdin2016,klatzow2019}, squeezed baths \cite{Rossnagel2014,manzano2016,klaers2017} or measurement channels \cite{elouard2017,elouard2018,Buffoni2019,Elouard2019} that could induce advantages to standard thermal machines.

Specifically, the role of measurement in relation to thermodynamics and information flow has been studied rigorously. For example, models of thermal machines facilitated by Maxwell's demon -- an external agent that acquires information of the system and performs appropriate feedback -- have been proposed in order to provide accurate thermodynamic description of information flow \cite{mandal2012,barato2013,strasberg2013,horowitz2014,Koski2015,Cottet2017,Koski2018}. More recently, a measurement channel has been deemed a source of ``quantum heat'' \cite{elouard2017} due to the increased entropy following a measurement, which could be exploited for both cooling \cite{Buffoni2019} and work extraction \cite{elouard2017maxwell,elouard2018,Elouard2019,Aydin2019}. However, proper treatment of actual erasure cost of pointers \cite{alicki2013,faist2015,alicki2019} as well as the interpretation of incoherent measurement schemes as a form of heat \emph{and} work exchange \cite{Strasberg2017,strasberg2018,seah2019} still remain a contentious topic for such measurement-based thermal machines. 

In this paper, we reveal the mechanisms underlying Maxwell's demon by considering a self-contained engine built from the standard ingredients (hot and cold reservoirs and a working medium) as well as an embedded macroscopic pointer. Specifically, we revisit definitions of work, heat, and information flow in a practical measurement-feedback scheme. In contrast to regular Maxwell-demon type engines where the demon has access to the state of the working medium and stores it in its memory, we restrict our demon's access to the pointer only, modelled by a damped mechanical degree of freedom. Work can then be extracted from the medium by reading off the pointer position and applying appropriate feedback. We show that such a setup generates a new type of engine with features different from standard quantum engines. In particular, we see that 
it is possible to attain simultaneous high powers and efficiencies based on the model's benchmarks. The regime of operation is also wider than that of a quantum Otto engine. 

\begin{figure}
    \centering
    \includegraphics[width=0.96\columnwidth]{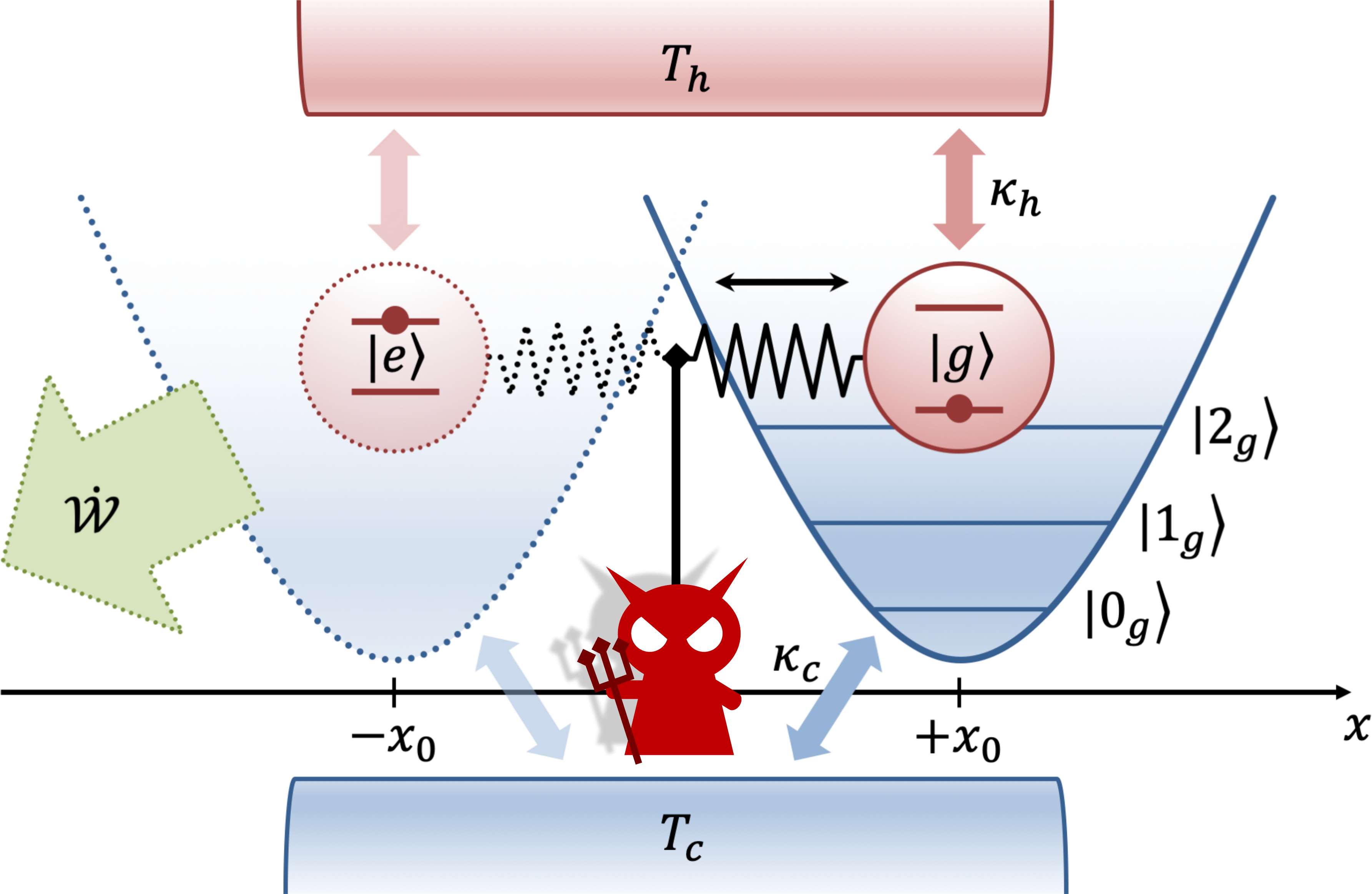}
    \caption{(Color online) Sketch of the demon system consisting of a qubit (working medium) and a harmonic oscillator (pointer). The qubit can be thermally excited by a hot bath at the rate $\kappa_h$ and temperature $T_h$, and it displaces the equilibrium position of the pointer to $\pm x_0$ depending on its state. A cold bath of temperature $T_c$ thermalizes the pointer around its equilibrium point at the rate $\kappa_c$. Work can be extracted coherently or incoherently from the excited spin by the demon's interrogation of the pointer position.}
    \label{fig:sketch}
\end{figure}

\emph{Spin-boson model.---} We consider a qubit with bare transition frequency $\Omega$ representing the working medium for heat-to-work conversion. A harmonic oscillator pointer of frequency $\omega$ couples to the qubit and is displaced to the left or right depending on the internal state of the qubit, see Fig.~\ref{fig:sketch}. The model Hamiltonian reads as
\begin{eqnarray}
    \oH &=& \frac{\hbar\Omega}{2} \osigma_z + \hbar\omega \left( \oa^\da \oa + \frac{1}{2}\right) + \hbar \omega x_0 \osigma_z \frac{\oa + \oa^\da}{\sqrt{2}} \nonumber \\
    &=& \frac{\hbar\Omega}{2} \osigma_z + \hbar\omega \ob^\da\ob + \text{const}, \label{eq:H}
\end{eqnarray}
with $\osigma_z = |e\ra\la e| - |g\ra\la g|$, $\oa$ the oscillator's mode operator and $\ob = \oa + \osigma_z x_0/\sqrt{2}$ the displaced mode operator. The Hamiltonian is found to be diagonal in the basis of qubit state-dependent displaced Fock states,
\begin{eqnarray}
    |g,n_g \ra := |g\ra \otimes \oD |n\ra,\quad |e,n_e \ra := |e\ra \otimes \oD^\dg |n\ra , \label{eq:basis} 
\end{eqnarray}
where the energy eigenvalues are $E_n^{e,g} = \pm \hbar\Omega/2 + \hbar\omega n$ modulo a constant, and $\oD = \exp\left(x_0 \oa^\da/\sqrt{2} - x_0\oa/\sqrt{2}\right)$ is the displacement operator.

A hot thermal reservoir with mean occupation number $\bar{n}_h= 1/\left[\exp(\hbar\Omega/k_BT_h)-1\right]$ injects heat and randomly excites the qubit, as mediated by the dissipators
\begin{eqnarray}
    \cL_h \rho &=& \sum_k \kappa_h (\Omega+k\omega) \Bigg \{ \left[ \bar{n}_h (\Omega+k\omega)+1\right] \label{eq:Lh_glo} \\
    && \times  \cD \left[ \sum_n d_{n,-k}^{*} |g,(n-k)_g\ra \la e,n_e| \right] \rho \nonumber \\
    &&+ \left. \bar{n}_h (\Omega+k\omega) \cD \left[ \sum_n d_{n,k} |e,(n+k)_e\ra \la g,n_g| \right] \rho \right\}, \nonumber 
\end{eqnarray}
with $\cD[\oA]\rho = \oA\rho\oA^\da - \{\oA^\da\oA,\rho\}/2$ and coefficients $d_{n,k} = \bra{n}\oD^2\ket{n+k}$. We derive \eqref{eq:Lh_glo} from a secular approximation of the weak coupling master equation (see Appendix).

A cold reservoir with $\bar{n}_c= 1/\left[\exp(\hbar\Omega/k_BT_c)-1\right]$  continuously couples to the pointer to erase/reset the information encoded in it. We employ thermal dissipators acting on the displaced mode operator $\ob$ \footnote{Eq.~\eqref{eq:Lc_glob} can be obtained from the usual Born-Markov secular approximation \cite{Breuer2002,Alicki2019a}, assuming an oscillator bath linearly coupled to the $\ox$-quadrature. This would also result in an additional pure dephasing term $\propto \cD[\osigma_z]\rho$, which however scales with the bath spectral density at zero frequency and is thus often negligible.},
\begin{equation}
    \cL_c \rho = \kappa_c (\bar{n}_c+1) \cD[\ob] \rho + \kappa_c \bar{n}_c \cD[\ob^\da] \rho. \label{eq:Lc_glob}
\end{equation}
Before introducing a demon for measurement-feedback, let us discuss the operation regime for this engine. Ideally, we want to work in the limit $\Omega \gg \omega \gg \kappa_c \gg \kappa_h$, which describes a separation of energy scales between the working medium and the pointer in the regime of resolved sidebands and weak thermal couplings. The pointer does not contribute appreciably to the energy balance ($\Omega \gg \omega$), but it reacts quickly to any change in the qubit state ($\kappa_c \gg \kappa_h$). Moreover, we require  sufficiently large $x_0$ compared to the thermal width $x_{\rm th} = \coth^{1/2}{\hbar\omega/2k_BT_c}\geq 1$ of the pointer, so that the pointer states become ``macroscopically distinguishable'' through their spatial separation \cite{alicki2013,alicki2019}. This corresponds to  ultrastrong qubit-oscillator coupling ($x_0>1$).

In the envisaged regime, the overall time evolution governed by Eqs.~\eqref{eq:H}-\eqref{eq:Lc_glob} brings the system to an approximate mixed steady state of the form
\begin{equation}
    \rho_\infty \approx (1-p_\infty) |g\ra\la g| \otimes \oD \rho_g \oD^\da + p_\infty |e\ra\la e| \otimes \oD^\da \rho_e \oD .
    \label{eq:SS_ideal}
\end{equation}
In particular, 
we have $p_\infty \approx \bar{n}_h/(2\bar{n}_h+1)$ and $\rho_{e,g} \approx  \exp(-\hbar\omega \oa^\da \oa/k_B T_c)/Z_c$ to lowest order in $\kappa_h/\kappa_c$, i.e.~a $T_h$-thermal mixture of displaced $T_c$-thermal pointer states encoding the qubit state. 
The demon will access the pointer position and perform conditioned feedback operations to extract energy from the qubit, and it will be functional so long as it possesses the ability to resolve the separated pointer states. This is unlike the case of a finite-dimensional pointer (e.g.~a qubit), whose states would not remain distinguishable in the presence of noise. Furthermore, if the demon were able to measure a qubit, it could measure the system directly and the pointer would be redundant \cite{sewell2007}.

We remark that our setup incorporates the practical cost of resetting the measurement apparatus: when the pointer reacts to a change in the qubit state and moves towards its new equilibrium point, the energy expelled to the cold bath amounts to $2\hbar\omega x_0^2\geq2\hbar\omega x_\mathrm{th}^2 > 4 k_B T_c$. This is always greater than the energy loss $k_B T_c \ln 2$ of an ideal Landauer erasure protocol.

Having set the model, we now introduce two demon configurations for work extraction: (1) an active agent performing random measurement-feedback and (2) a passive agent in the form of a coherent control field continuously monitoring the pointer.

\emph{Active demon.---}
We first consider an active demon that interrogates the pointer position and performs necessary feedback at a rate $\gamma$ based on the following protocol:
(i) a dichotomic projective measurement ($\oP$ and $1-\oP$) of the pointer to detect whether it is on the left ($\la \ox\ra <0$)\footnote{An ideal position measurement is such that $\oP_x = \int_{-\infty}^0 \dd x\, \ketbra{x}{x}$. For numerical simulations, we consider instead $\oP = \sum_{n=0}^N \oD^\da \ketbra{n}{n}\oD$ for numerical stability while avoiding convergence issues due to discontinuities in position representation. Here, the cutoff $N$ is chosen such that the included pointer levels contain at least the ground state, but otherwise do not exceed the potential energy of the left-displaced oscillator at $x=0$, i.e.~$2N+1 \leq \max \{x_0^2,1\}$.}, followed by (ii) work extraction via a Rabi flip $\osigma_x$ induced by a strong control pulse if the pointer is on the left, i.e.~the qubit is most probably excited. 
Notice that the measurement step (i) induces transitions between the energy eigenstates of the pointer due to measurement backaction, since $[\oH,\oP] \neq 0$. While this can be interpreted as a form of ``quantum heat'' \cite{elouard2017}, the net energy change will be small compared to the extraction step (ii).
For infinitesimally short and sufficiently sparse Poisson-distributed events, the process can be effectively described by the coarse-grained generator \cite{Breuer2002,Hornberger2007,wiseman2009,Jacobs2014}
\begin{equation}
    \cL_m \rho = \gamma \cD[\osigma_x \oP ] \rho + \gamma \cD[\oP]\rho, \label{eq:Lm}
\end{equation}
which leads to a minor perturbation of the steady state $\rho_\infty$ as long as $\gamma \ll \kappa_c$.

Assuming vanishing overlap between the two displaced pointer states such that \eqref{eq:SS_ideal} is reached, the demon would ideally generate a maximum energy output of $W_{\max} = \hbar(\Omega-2\omega x_0^2)p_\infty $ by application of a spin flip on \eqref{eq:SS_ideal}.
In fact, such an intuitive scheme is sufficient for extracting energy close to the ergotropy \cite{allah2004work} (maximum extractable energy from a quantum system by means of a cyclic unitary transformation) contained in \eqref{eq:SS_ideal}, $W_{\rm erg} \approx \hbar(\Omega-\omega) p_\infty - \hbar\omega \bar{n}_c $ for $\bar{n}_h > \bar{n}_c$.

The present scheme does not rely on externally imposed engine strokes with synchronized switching of control pulses or couplings to thermal reservoirs. The random measurement process not only facilitates a convenient assessment of stationary energy flows,  $\dot{Q}_{c,h,m} = \tr \{ \oH \cL_{c,h,m} \rho_\infty \}$, but it also does not depend on the precise timing of ``measurement strokes''. 
Specifically, the steady-state power due to $\cL_m$ consists of two terms,
$\dot{Q}_m = \dot{Q}_{\rm ba} - \dot{W}$, with 
\begin{eqnarray}
    \dot{Q}_{\rm ba} &=& 2\gamma \tr \left\{ \oH \cD[\oP]\rho_\infty \right\} = 2 \gamma \hbar\omega \, \tr \left\{ \ob^\da\ob \cD[\oP]\rho_\infty \right\}, \nonumber \\
    \dot{W} &=& \gamma \tr \left\{ \oP\rho_\infty \oP \left[ \oH - \osigma_x \oH \osigma_x \right] \right\} \nonumber \\ 
    &=& \gamma \tr \left\{ \oP\rho_\infty \oP \left[ \hbar\Omega + 2\hbar\omega x_0 \ox \right]\osigma_z \right\}. \label{eq:Wincoh}
\end{eqnarray}
Here, $\dot{Q}_{\rm ba}$ describes the pure backaction effect of pointer measurement without feedback coming from a unital channel that increases the system's entropy, and this energy would have to come from the source implementing the projectors. Meanwhile, $\dot{W}$ stems from the $\osigma_x$-feedback and can be understood as the average rate of \emph{useful} energy extracted by performing a spin flip on the post-measurement state $\oP \rho_{\infty}\oP$.

\begin{figure} 
    \centering
    \includegraphics[width=\columnwidth]{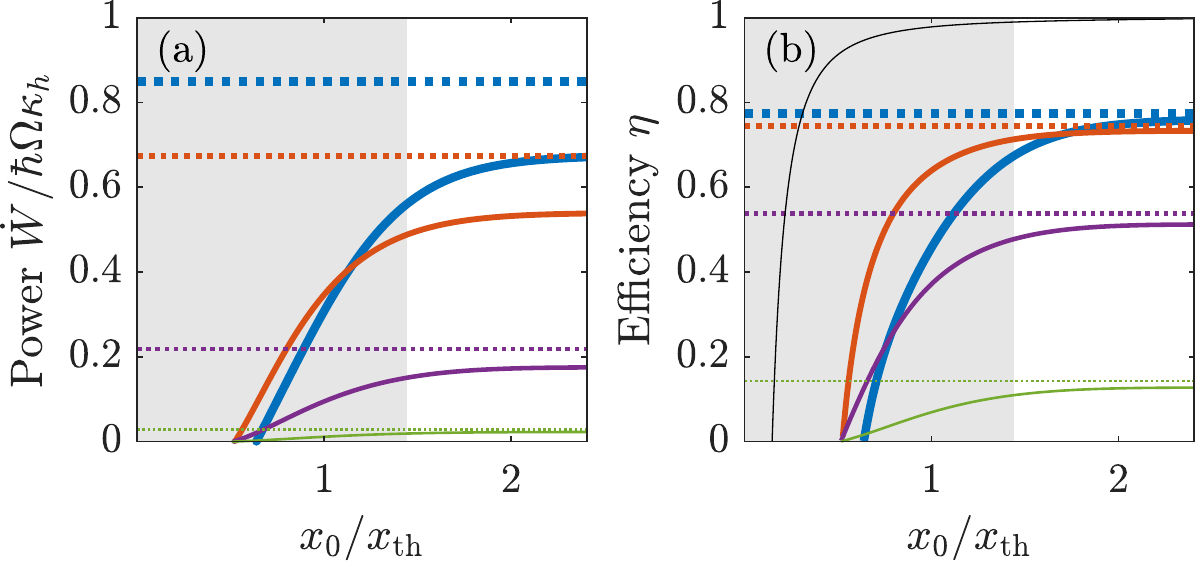}
    \caption{(Color online) Steady-state output power (a) and efficiency (b) as a function of $x_0/x_{th}=x_0\sqrt{\tanh(\hbar\omega/2k_BT_c)}$ ($T_c=0$ on the right). The blue (thick), red, purple, and green (thin) curves correspond to measurement rates $\gamma/\omega = 10^{-1}$, $10^{-2}$, $10^{-3}$, and $10^{-4}$, respectively. The horizontal dotted lines represent the approximations $\gamma W_{\max}$ and \eqref{eq:effincoh}, the shaded regions mark $\bar{n}_c \geq \bar{n}_h$, while the black solid line in (b) shows the Carnot efficiency $\eta_{\rm Carnot}=1-T_c/T_h$. The Otto efficiency $\eta_{\rm Otto}=1-\omega/\Omega=0.99$ is constant. We fix $\Omega=100 \omega$, $x_0=2.5$, $\kappa_h = 10^{-3}\omega$, $\kappa_c=0.1\omega$, $\bar{n}_h = 1$ i.e.~$T_h=\hbar\Omega/k_B\ln 2$. }
    \label{fig:Wincoh_T}
\end{figure}

When the measurement rate $\gamma \ll \kappa_c$ and the pointer separation $x_0 \gg 1$, the projector would reduce the state to the excited branch in \eqref{eq:SS_ideal} resulting in the benchmark power $\gamma W_{\max}$. The repeated measurements however diminish the branch weight, $p_\infty \approx \bar{n}_h/(2\bar{n}_h+1+\gamma/\kappa_h)$ (see Appendix).  For the efficiency, $\eta = \dot{W}/\dot{Q}_h$, we find the approximate upper bound\footnote{Depending on interpretation, one could include the cost of the measurement process in the efficiency, either as an additional ``heat'' input, $\eta = \dot{W}/(\dot{Q}_h + \dot{Q}_{\rm ba})$, or as work expense reducing the net output, $\eta = (\dot{W} - \dot{Q}_{\rm ba})/\dot{Q}_h$. The definitions coalesce in the regime of operation considered, since $\dot{Q}_{\rm ba}\propto \omega$ whereas $\dot{W},\dot{Q}_h\propto\Omega\gg\omega$.}
\begin{equation}
    \eta_{\max} \approx \frac{1-2\omega x_0^2/\Omega}{1+ 2[1+(2\bar{n}_h+2)\kappa_h/\gamma]\omega x_0^2/\Omega} .\label{eq:effincoh}
\end{equation}
Both the output power and efficiency grow with $\gamma$ until an optimum is reached around $\gamma \lesssim \kappa_c$. At higher $\gamma$, we eventually reach a Zeno limit where frequent measurements hinder the pointer from moving between the left and the right equilibrium, essentially freezing the engine operation.

Figure \ref{fig:Wincoh_T} shows (a) the output powers and (b) efficiencies as a function of $T_c$ for various rates $\gamma$. Here, $T_c$ is expressed in terms of the ratio between pointer displacement $x_0$ and characteristic thermal width $x_{\rm th}$. This is an exemplary case where $x_0=2.5$, which should lead to a clear separation of the ground- and excited-state distributions so long as the cold bath temperature is sufficiently low ($x_0>x_{\rm th}$). As our demon scheme captures the measurement and erasure costs through a mechanical pointer continuously reset by the cold bath, the engine operation is consistent with the second law of thermodynamics and the efficiencies do not exceed the Carnot bound.

In the low-$T_c$ limit, the efficiencies and output powers approach the analytical benchmarks given by \eqref{eq:effincoh} and $\gamma W_{\rm max}$ respectively, especially for small $\gamma$ where the measurement effect is negligible and the steady state can by approximated by \eqref{eq:SS_ideal}. 
At high $T_c$, the efficiencies and powers fall below the benchmark and the output power eventually becomes negative due to the larger overlap between the two displaced thermal states, which leads to inaccurate readout of the qubit state.

Should the macroscopic pointer be replaced with a qubit, the operation would be restricted to the standard Otto window ($\bar{n}_h>\bar{n}_c$). This is because feedback errors leading to work consumption instead of extraction would proliferate with growing $\bar{n}_c$ and the net work output per interrogation would be limited by $\hbar (\Omega-\omega)(\bar{n}_h - \bar{n}_c)/(2\bar{n}_h+1)(2\bar{n}_c+1)$ and lead to an Otto efficiency $\eta_{\rm Otto}=1-\omega/\Omega$ \cite{lloyd1997}. In our model with a macroscopic pointer, we see that the engine operates well beyond the Otto window (shaded region in Fig.~\ref{fig:Wincoh_T}) so long as the pointer states are spatially distinguishable, i.e.~when $x_0\gtrsim x_{\rm th}$. At vanishing $\kappa_h$, the system reaches a maximum efficiency that is lower than Otto, $\eta \approx 1-4 x_0^2 \omega/\Omega < \eta_{\rm Otto}$. It can be attained simultaneously with the maximum power $\gamma W_{\rm max}$.

\emph{Passive demon.---} Instead of an incoherent scheme based on random monitoring by an external agent, it would be insightful to formulate an integrated setup in which the measurement-feedback takes place internally and all energy exchanges become transparent: we do not have to deal with work cost associated to $\dot{Q}_{\rm ba}$. To this end, we consider a position-dependent driving field of strength $\zeta$ with detuning $\Delta$, which now plays the role of the demon that probes the qubit in a non-invasive, coherent manner. We can describe the effect of such a demon by a time-dependent Rabi term
\begin{equation}
    \oV (t) = \hbar \zeta f(\ox) e^{-i(\Omega-\Delta)t} |e\ra \la g| + h.c. . \label{eq:Vdrive}
\end{equation}
The field serves as an interface for continuous work extraction depending on the position-dependent function $f(\ox)$, bearing similarities to work extraction via coherent pulses from a cyclic demon engine previously considered in \cite{Cottet2017}. Possible choices of $f(x)$ include a Heaviside function $\Theta(-x)$ or a Gaussian centred around $x = -x_0$. 

To assess the scheme's steady-state performance, we consider the weak driving limit, $\zeta \ll \omega,\Omega$, where corrections to the thermal dissipators $\cL_{h,c}$ can be omitted \cite{carmichael2003,Szczygielski2013}. In the frame rotating at the driving frequency, the time dependence due to \eqref{eq:Vdrive} conveniently disappears and the time evolution follows from $\tilde{\cL}_{h,c} = \cL_{h,c}$ and $\otH/\hbar = \Delta \osigma_z/2 + \omega \ob^\da\ob +\zeta f(\ox)\osigma_x$. The corresponding steady state $\tilde{\rho}_\infty$ describes the engine's limit cycle and yields the average output power \cite{alicki1979quantum}
\begin{equation}
    \dot{W} = -\tr \left\{ \rho_\infty (t) \partial_t \oV(t) \right\} = -\hbar\zeta (\Omega-\Delta) \tr \left\{ f(\ox) \osigma_y \tilde{\rho}_\infty \right\}. \label{eq:Wcoh}
\end{equation}
The heat fluxes from the hot and cold reservoirs read as
\begin{equation}
    \dot{Q}_{h,c} = \tr \left\{ \left[ \otH + \hbar\frac{\Omega - \Delta}{2}\osigma_z \right] \cL_{h,c} \tilde{\rho}_\infty \right\}. \label{eq:Qcoh}
\end{equation}

\begin{figure}
    \centering
    \includegraphics[width=\columnwidth]{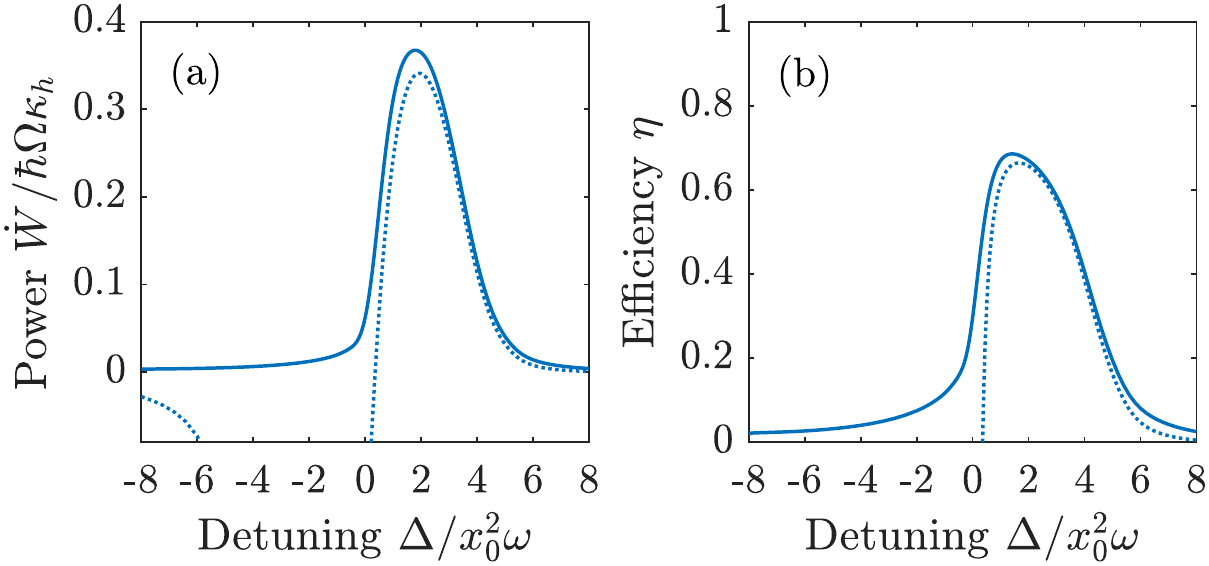}
    \caption{(Color online) Output power (a) and efficiency (b) against the detuning $\Delta$ of the driving field for $f(x) = \Theta(-x)$ (solid) and $f(x) = 1$ (dotted). We fix $\Omega=100 \omega$, $x_0=2.5$, $\kappa_h = 10^{-3}\omega$, $\kappa_c=0.1\omega$, $\zeta = 0.1\omega$ and the same hot and cold bath occupancy $\bar{n}_c = \bar{n}_h = 1$. The Carnot and Otto efficiencies are both 0.99.}
    \label{fig:Wcoh_detuning}
\end{figure}

\begin{figure}
    \centering
    \includegraphics[width=\columnwidth]{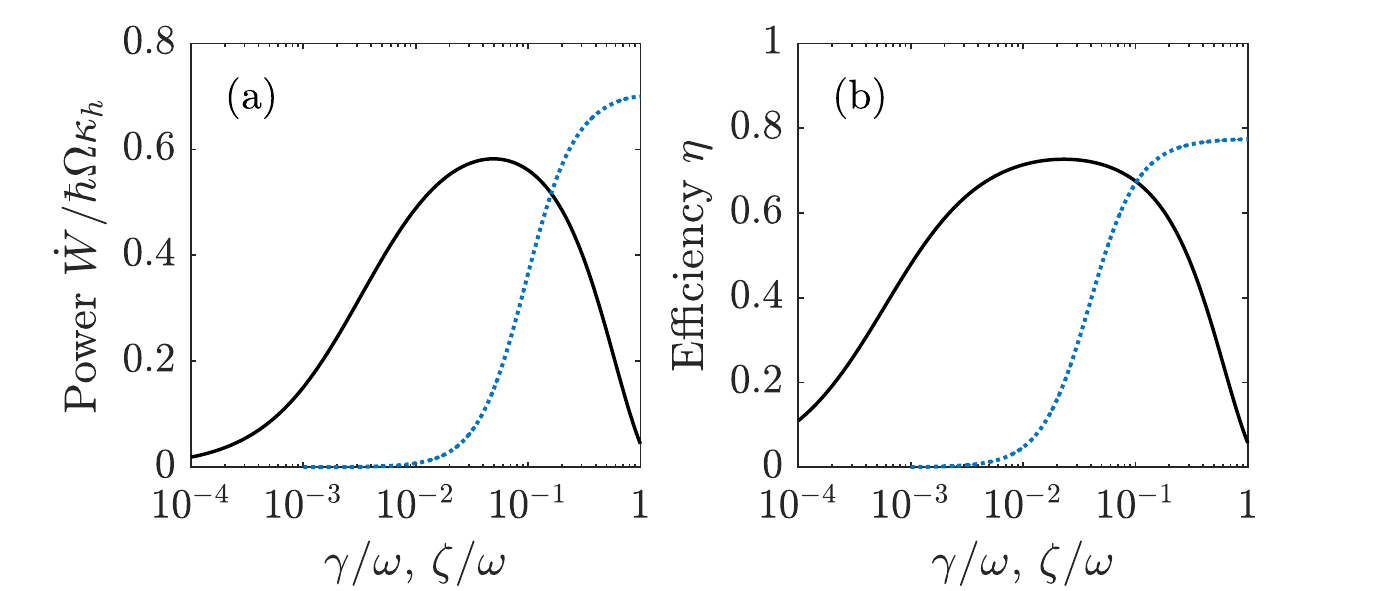}
    \caption{(Color online) Output power (a) and efficiency (b) against the rate $\gamma$ or $\zeta$, comparing the active measurement-feedback scheme (black solid) with the passive scheme (blue dotted) at optimal detuning $\Delta=2\omega x_0^2$ and $f(x) = \Theta(-x)$. The other parameters are taken from Fig.~\ref{fig:Wcoh_detuning}. Note that underlying master equation model may no longer be reliable for $\zeta \sim \omega$. }
    \label{fig:cohVSincoh}
\end{figure}

Figure \ref{fig:Wcoh_detuning} shows the engine's output powers and efficiencies at its limit cycle as a function of the detuning for an exemplary set of engine parameters and various cold bath temperatures. Here, the optimal output power is much smaller than the driving rate times the extractable excitation energy, $\zeta W_{\max} \approx 29 \hbar\Omega \kappa_h$. This was not the case for the previously discussed incoherent measurement-feedback scheme, which exhibits a work power of up to $\gamma W_{\max}$, because that scheme implicitly assumes a large driving strength and short feedback time such that the feedback is essentially described by a conditional spin flip depending on the position of the pointer. In the current scheme, the driving field would not cause a full spin flip. Nevertheless, the output power can be comparable to what the measurement-feedback scheme predicts for similar settings, see also Fig.~\ref{fig:cohVSincoh}.

Here, we achieve a maximum work power (and efficiency) when $\Delta \approx 2\omega x_0^2$. This is because the frequency of the qubit is modulated by the pointer position, and at this driving frequency, the field addresses predominantly the qubit only when the pointer is located at $-x_0$, i.e.~the qubit is excited and the field is able to extract a positive net energy from it. Hence one can modify the scheme by removing the position dependence $f(x)$ and consider a non-invasive interrogation of the qubit state solely through the application of a red-detuned field of $\Delta \approx 2\omega x_0^2$. This does not cause a backaction-induced direct flow of energy to the pointer, a minor contribution to the energy balance when $\Omega \gg \omega$, which is inherent to the position-dependent case and appears explicitly as $\dot{Q}_{\rm ba}$ in the previous measurement-feedback scheme.

The dotted line in Fig.~\ref{fig:Wcoh_detuning} shows the output power and efficiency achievable by non-invasive interrogation as a function of the detuning. Close to the optimal working point, the performance is almost the same as the position-dependent case, but the position-independent driving will cease to produce work as the detuning approaches zero; indeed, we would obtain a heat pump \emph{consuming} work at negative detunings. 

Finally, Figure \ref{fig:cohVSincoh} compares the active and passive demon at optimal detuning and position-dependent driving in terms of their  powers and efficiencies. We plot them as a function of the respective interrogation rates $\gamma$ and $\zeta$. The active scheme performs well over a broad range of small measurement rates $\gamma$, but it stops working when the Zeno effect kicks in at $\gamma > \kappa_c$. The passive scheme eventually catches up at strong driving rates $\zeta$.

\emph{Experimental platforms.---} Regarding implementations, the proposed Hamiltonian \eqref{eq:H} would describe molecular batteries \cite{Alicki2019a}: molecules with an optical electronic transition strongly coupled to an infrared vibration mode. It also resembles the Holstein Hamiltonian for a molecule undergoing fast vibrational relaxation \cite{Reitz2019}, where displacements can reach magnitudes $x_0 \sim 1$, while the vibrational relaxation time is short compared to the optical lifetime, i.e.~$\kappa_h \ll \kappa_c$. A broadband optical light source (e.g.~filtered sunlight) could serve as the hot bath exciting the electron, and a resonant IR cavity mode could be employed to monitor the vibration mode displacement \cite{Long2015,Shalabney2015}. 
Alternatively, hybrid optomechanical systems would be a natural platform to incorporate a macroscopic pointer in the ultrastrong regime \cite{yeo2014,Treutlein2014,Monsel2018}. Our scheme could also be realized in a tailored trapped-ion setup similar to the recently demonstrated spin-flywheel engine \cite{Lindenfels2019}.

\emph{Conclusions.---}
We presented a self-contained engine model in which useful energy is extracted from thermal excitations of a quantum spin by a restricted demon that can only interrogate the spin state through the position of a macroscopic pointer attached to the spin. Our work reveals the fundamental energy fluxes for an autonomous Maxwell's demon engine including work extraction, measurement backaction and information transfer. Specifically, we evaluated the engine performance both for an active demon performing measurement-feedback events at random times and for a passive demon in the form of a stationary control field. While the use of a macroscopic pointer shows that the energy loss associated with erasure/reset would exceed Landauer erasure in reality, it also allows the engine to operate beyond typical operation windows in quantum engines, putting forth the paradigm of continuous measurement-driven engines. 

\emph{Acknowledgments.---}The authors acknowledge fruitful discussions with Robert Alicki and Claudiu Genes. This research is supported by the National Research Foundation and the Ministry of Education, Singapore, under the Research Centres of Excellence programme.

\clearpage
\begin{widetext}

\appendix

\section*{Appendix: Validity of the hot bath dissipator}\label{app:hotlocalVSglobal}

Here we compare the local and global secular form of the hot bath dissipator that arises in the usual manner from a linear exchange interaction of the qubit with a thermal oscillator bath. For the local model, one simply employs the standard dissipator for an isolated qubit,
\begin{equation}
    \cL_h^{\rm loc} \rho = \kappa_h (\Omega) [\bar{n}_h(\Omega) +1] \cD[\osigma_-]\rho + \kappa_h (\Omega) \bar{n}_h(\Omega) \cD[\osigma_+]\rho,
\end{equation}
assuming that the qubit-pointer coupling and thus the influence of the pointer on the qubit energy are negligible. 
For an isolated qubit, the jump operators $\osigma_{\pm}$ can mediate only a single transition of frequency $\Omega$, given the thermal coupling rate $\kappa_h$ and the mean thermal bath occupation $\bar{n}_h$ at this frequency.
In the combined qubit-pointer system, the same operators now induce a family of transitions $\Omega + k\omega$ with $k\in \mathbb{Z}$. Specifically, we can expand in terms of the combined energy basis \eqref{eq:basis},
\begin{equation}
    \osigma_+ = |e\ra\la g| = \sum_{m,n=0}^\infty \la m|\oD^2 |n\ra |e,m_e\ra \la g,n_g| = \sum_{n=0}^\infty \sum_{k=-n}^\infty  \underbrace{\la n+k|\oD^2 |n\ra}_{\equiv d_{n,k}} |e,(n+k)_e\ra \la g,n_g|,
\end{equation}
with the weight coefficients $d_{n,k}$. 
The above local dissipator contains cross-terms between different transitions $k\neq k'$, which means that it preserves a certain amount of coherences between different energy levels of the system. Moreover, using it implies that one can neglect the frequency dependence of the bath parameters, $\kappa_h(\Omega + k\omega) \approx \kappa_h$ and $\bar{n}_h(\Omega + k\omega) \approx \bar{n}_h$, which is only valid when $\Omega \gg \omega$. 

The global secular model does not preserve any coherences between different Fock numbers, because it contains only resonant jump terms,
\begin{eqnarray}
    \cL_h^{\rm glo} \rho &=& \sum_k \kappa_h (\Omega+k\omega) \left[ \bar{n}_h (\Omega+k\omega)+1\right] \cD \left[ \sum_n d_{n,-k}^{*} |g,(n-k)_g\ra \la e,n_e| \right] \rho \nonumber \\
    &&+ \sum_k \kappa_h (\Omega+k\omega) \bar{n}_h (\Omega+k\omega) \cD \left[ \sum_n d_{n,k} |e,(n+k)_e\ra \la g,n_g| \right] \rho.
\end{eqnarray}
For the demon models studied in the main text, we find that both dissipators yield approximately the same results. The reason is, on the one hand, that we indeed consider $\Omega \gg \omega$ and can thus assume constant $\kappa_h$ and $\bar{n}_h$. On the other hand, our model also includes a cold bath with stronger damping rate $\kappa_c > \kappa_h$, which suppresses any coherences between Fock states of the pointer that $\cL_h^{\rm loc}$ alone would have preserved. 

The steady-state heat input for $\kappa_h \ll \kappa_c$ can then be approximated using the local dissipator, too,
\begin{equation}
    \dot{Q}_h \approx \tr \left\{ \left( \frac{\hbar \Omega}{2} + \hbar\omega x_0 \ox \right) \osigma_z \cL_h^{\rm loc} \rho_\infty \right\} \approx \hbar \kappa_h \left[ \bar{n}_h (1-p_\infty)(\Omega+2\omega x_0^2) - (\bar{n}_h+1)p_\infty (\Omega-2\omega x_0^2) \right].
\end{equation}
For the qubit excitation probability, the same approximation yields
\begin{eqnarray}
    \partial_t p_e (t) &\approx& \tr \left\{ |e\ra\la e| (\cL_h^{\rm loc} + \cL_m)\rho \right\} = -\kappa_h (\bar{n}_h+1) p_e(t) + \kappa_h \bar{n}_h [1-p_e(t)] - \gamma \tr \left\{ \osigma_z \oP\rho\oP \right\} \nonumber \\
    &\approx& -\kappa_h (\bar{n}_h+1) p_e(t) + \kappa_h \bar{n}_h [1-p_e(t)] - \gamma p_e(t).
\end{eqnarray}
Here the second line holds in the ideal operation regime of $\gamma \ll \kappa_c$ and $x_0 \gg 1$, when $\oP$ reduces the state to its excited branch. At steady state, we obtain $p_\infty = p_e(\infty) = \bar{n}_h/(2\bar{n}_h + 1 + \gamma/\kappa_h)$, as used in the main text. 

\end{widetext}

\end{document}